\documentclass[aps,prb,twocolumn,showpacs,superscriptaddress]{revtex4}%
\usepackage{amsfonts}
\usepackage{amsmath}
\usepackage{amssymb}
\usepackage{amsbsy}
\usepackage{graphicx}%
\providecommand{\U}[1]{\protect\rule{.1in}{.1in}}

\begin{document}
\title{Intrinsic spin-relaxation induced negative tunnel magnetoresistance in a
single-molecule magnet}
\author{Haiqing Xie}
\affiliation{Institute of Theoretical Physics and Department of Physics, Shanxi University,
Taiyuan 030006, China}
\author{Qiang Wang}
\affiliation{Institute of Theoretical Physics and Department of Physics, Shanxi University,
Taiyuan 030006, China}
\author{Hai-Bin Xue}
\affiliation{College of Physics and Optoelectrics, Taiyuan University of Technology,
Taiyuan 030024, China}
\author{HuJun Jiao}
\affiliation{Institute of Theoretical Physics and Department of Physics, Shanxi University,
Taiyuan 030006, China}
\author{J.-Q. Liang}
\email{jqliang@sxu.edu.cn}
\affiliation{Institute of Theoretical Physics and Department of Physics, Shanxi University,
Taiyuan 030006, China}
\keywords{Spin polarized transport; single-molecule magnet}
\pacs{75.50.Xx, 75.76.+j, 85.65.+h, 85.75.d}

\begin{abstract}
We investigate theoretically the effects of intrinsic spin-relaxation on the
spin-dependent transport through a single-molecule magnet (SMM), which is
weakly coupled to ferromagnetic leads. The tunnel magnetoresistance (TMR) is
obtained by means of the rate-equation approach including not only the
sequential but also the cotunneling processes. It is shown that the TMR is
strongly suppressed by the fast spin-relaxation in the sequential region and
can vary from a large positive to slight negative value in the cotunneling
region. Moreover, with an external magnetic field along the easy-axis of SMM,
a large negative TMR is found when the relaxation strength increases. Finally,
in the high bias voltage limit the TMR for the negative bias is slightly
larger than its characteristic value of the sequential region, however it can
become negative for the positive bias caused by the fast spin-relaxation.

\end{abstract}
\date{\today}
\maketitle


\section{Introduction}

The electron transport through magnetic molecules has attracted much attention
recently in molecular spintronics
\cite{Wernsdorfer,Timm1,Kim,exp,Berry,Kondo,Martin1,Martin2,JB,JBrel,Timm2,Rossier,
Shen,Xing,Xie,RWang,JBdiode,TimmM,JBTMR1,JBTMR2,JBTMR3,TAMR,Timmco,Timm3,exp1,Sci,Sothmann}%
, because of its potential applications in information storage and processing.
The single-molecule magnet (SMM), which has a high intrinsic spin and an
easy-axis magnetic anisotropy, is of particular interest. Some peculiar
phenomena such as Coulomb-blockade, magnetic excitations, complete
current-suppression, and negative differential conductance have been observed
experimentally in the quantum transport through a SMM \cite{exp}. The Berry
phase \cite{Berry} and Kondo effects \cite{Kondo} on the average current (or
differential conductance) are predicted theoretically. Moreover, the full
counting statistics (current fluctuation) \cite{Martin1} and spin fluctuation
of magnetic molecules \cite{Martin2} are also studied. On the other hand, the
magnetic switching of SMM is realized by thermal spin-transfer torque
\cite{Xing}, spin-bias \cite{Shen}, and spin-polarized current injected from
ferromagnetic (FM) electrodes \cite{Martin2,JB,JBrel,Timm2,Rossier}. It is
predicted that the spin-current polarization can be reversed through a SMM
with FM leads \cite{Xie} and even a pure spin-current can be generated in
normal leads by the thermoelectric effects \cite{RWang}. Spin-diode behavior
\cite{JBdiode} and memristive properties \cite{TimmM} are also observed in
spin-polarized transport through a SMM. The current difference between the
parallel (P) and antiparallel (AP) configurations of two FM electrodes known
as the tunnel magnetoresistance (TMR) \cite{Jullier} was systematically
investigated for the SMM by Misiorny \textit{et al.} in the sequential,
cotunneling, and Kondo regions \cite{JBTMR1,JBTMR2,JBTMR3}. The TMR is
positive usually, since the P current is greater than the AP one, while a
large negative TMR is shown to exist in the SMM \cite{JBTMR1}. Furthermore,
the tunneling anisotropic magnetoresistance (TAMR) with tunable magnitude and
sign is found in SMM junctions with only one FM electrode \cite{TAMR}.

A long spin-relaxation time of the SMM is required in the quantum information
processing \cite{Loss} and information storage \cite{Timm2}. However, in
addition to the spin-relaxation induced by the electron tunneling through
molecules, the SMM itself may have intrinsic spin-relaxation
\cite{JBrel,Timmco}, which originates from the interactions with environmental
spins. A short spin-relaxation time is found in a recent experiment
\cite{exp1} in which a spin-polarized scanning tunneling microscopy is used to
pump the spins of Mn atoms. Moreover, the experimentally observed quenching of
conductance steps \cite{Sci} has been successfully explained in terms of the
spin-relaxation \cite{Sothmann}. The effects of intrinsic spin-relaxation on
the current-induced magnetic switching of SMM have been studied theoretically
\cite{JBrel}. When the relaxation of SMM is very fast, the fine structure of
transport is suppressed, but a high spin-polarized current in nonmagnetic
leads can be generated with an external magnetic field \cite{Timmco}. For the
quantum-dot (QD) systems, on the other hand, the intrinsic spin-relaxation can
result in the TMR suppression \cite{JBQD1} and reversal \cite{JBQD2} in the
sequential and cotunneling regions, respectively. It is also shown that the
peak heights of differential conductance in QDs are increased by the
spin-relaxation \cite{QD}.

In this paper we study the spin-polarized transport through a SMM with FM
leads in both sequential and cotunneling regions and pay particular attention
on the intrinsic spin-relaxation, which affects remarkably the transport
properties. A large negative TMR can be obtained when a magnetic field is
applied along the easy-axis of SMM. Sec. II is devoted to the rate-equation
formalism for the spin-polarized transport through a SMM and the related TMR.
We demonstrate and discuss the effects of intrinsic spin-relaxation on the
current, differential conductance, magnetization and TMR based on the
numerical results in Sec.III. Finally, the conclusion is given in Sec. IV.

\begin{figure}[pth]
\includegraphics[width=0.8\columnwidth]{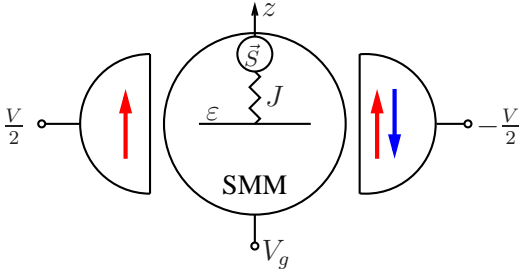}\caption{(Color online)
Schematic diagram of electron transport through a SMM weakly coupled to
ferromagnetic electrodes. The magnetizations of the two leads are collinear
with the magnetic easy axis of SMM (as z axis). The left (L) and right (R)
electrodes are connected with the bias voltage V/2 and -V/2, respectively.}%
\label{Fig:1}%
\end{figure}

\section{Model and method}

The schematic diagram of electron transport through a SMM coupled to two
external FM leads is shown in Fig. 1. It is assumed that the
lead-magnetizations are collinear with the magnetic easy-axis of SMM in either
P or AP configuration. The total Hamiltonian of the system can be written as
\cite{Timm2,Timm3,Timmco,JBTMR1}
\begin{equation}
H=H_{leads}+H_{SMM}+H_{T}.
\end{equation}
The first term $H_{leads}=\sum_{\mathbf{\alpha=L,R}}\sum_{\mathbf{k}\sigma
}\varepsilon_{\alpha\mathbf{k}}c_{\alpha\mathbf{k}\sigma}^{\dag}%
c_{\alpha\mathbf{k}\sigma}$ describes noninteracting electrons in electrodes,
where $c_{\alpha\mathbf{k}\sigma}^{\dag}$ ($c_{\alpha\mathbf{k}\sigma}$) is
the creation (annihilation) operator for an electron with wave vector
$\mathbf{k}$ and spin $\sigma$ in the lead $\alpha$, and $\varepsilon
_{\alpha\mathbf{k}\sigma}$ is the corresponding electron energy. The spin
polarization of FM lead $\alpha$ is defined as $p_{\alpha}=(\rho_{\alpha
+}-\rho_{\alpha-})/(\rho_{\alpha+}+\rho_{\alpha-})$, with $\rho_{\alpha+(-)}$
denoting the density of states for the majority (minority) electrons. The
second term$\ \ $
\begin{align}
H_{SMM}  &  =\sum_{\sigma}(\varepsilon-eV_{g})d_{\sigma}^{\dag}d_{\sigma
}+Ud_{\uparrow}^{\dag}d_{\uparrow}d_{\downarrow}^{\dag}d_{\downarrow
}-J\mathbf{s}\cdot\mathbf{S}\nonumber\\
&  -K(S^{z})^{2}-B(s^{z}+S^{z})
\end{align}
denotes the SMM Hamiltonian, where the operator $d_{\sigma}^{\dag}$
($d_{\sigma}$) creates (annihilates) an electron in the lowest unoccupied
molecular orbital (LUMO) level and $\mathbf{s}\mathbf{\equiv}$ $\sum
_{\sigma\sigma^{\prime}}d_{\sigma}^{\dag}(\mathbf{\sigma}_{\sigma
\sigma^{\prime}}/2)d_{\sigma^{\prime}}$ is the electron spin operator with
$\mathbf{\sigma}$ being the vector of Pauli matrices. The single-electron
energy $\varepsilon$ in the LUMO level is tunable by gate voltage $V_{g}$. The
Coulomb energy denoted by $U$ is for the double occupancy of the LUMO level.
The exchange coupling $J$ between the local spin of molecule $\mathbf{S}$ and
the LUMO electron-spin $\mathbf{s}$ can be either FM ($J>0$) or
antiferromagnetic (AFM) ($J<0$). The easy-axis anisotropy of the SMM is
characterized by the parameter $K$ ($K>0$) and an external magnetic field $B$
is applied along the easy-axis, where $B$ contains the factor $g\mu_{B}$.
Since the SMM Hamiltonian $H_{SMM}$ commutes with $z$ component $S_{t}^{z}$ of
the total spin operator $\mathbf{S}_{t}\mathbf{\equiv s+S}$, it can be easily
diagonalized in terms of molecular many-body eigenstates $\left\vert
n,S_{t};m\right\rangle $ with $n$ denoting the charge state, $S_{t}$ the total
spin quantum-number, and $m$ the eigenvalues of $S_{t}^{z}$.

The tunneling Hamiltonian between the LUMO level and leads is
\cite{Timm2,Timm3,Timmco,JBTMR1}
\begin{equation}
H_{T}=\sum_{\alpha\mathbf{k}\sigma}(t_{\alpha}c_{\alpha\mathbf{k}\sigma}%
^{\dag}d_{\sigma}+t_{\alpha}^{\ast}d_{\sigma}^{\dag}c_{\alpha\mathbf{k}\sigma
}),
\end{equation}
where the parameter $t_{\alpha}$ is the tunneling coupling-constant between
the molecule and lead $\alpha$, and\textbf{\ }the corresponding spin-dependent
tunnel-coupling strength is denoted by\textbf{ }$\Gamma_{\alpha\sigma}%
=2\pi\rho_{\alpha\sigma}\left\vert t_{\alpha}\right\vert ^{2}$\textbf{.}
Furthermore, in terms of the spin polarization $p_{\alpha}$ of lead $\alpha$,
we can rewrite the tunnel-coupling strength as $\Gamma_{\alpha\pm}%
=\Gamma_{\alpha}(1\pm p_{\alpha})/2$ for the spin-majority (spin-minority)
electrons of the lead $\alpha$, with $\Gamma_{\alpha}=\Gamma_{\alpha+}%
+\Gamma_{\alpha-}$.

Since the SMM-electrode coupling is assumed to be sufficiently weak, i.e.,
$\Gamma_{\alpha\sigma}\ll k_{B}T$, the molecule relaxes rapidly to the
eigenstates of $H_{SMM}$ (rapid dephasing) \cite{Timmco}. Therefore, the
rate-equation approach can be adopted to study spin-dependent transport
through the SMM. Here, we consider phenomenologically the intrinsic
spin-relaxation processes, which may drive the SMM to an eigenstate of lower
energy. The transport dynamics is well described by the following rate
equation,
\begin{align}
\frac{dP_{i}}{dt}  &  =\sum_{\alpha\alpha^{\prime}i^{\prime}\neq
i}[-(W_{\alpha}^{i,i^{\prime}}+W_{\alpha,\alpha^{\prime}}^{i,i^{\prime}%
}+W_{rel}^{i,i^{\prime}})P_{i}\nonumber\\
&  +(W_{\alpha}^{i^{\prime},i}+W_{\alpha^{\prime},\alpha}^{i^{\prime}%
,i}+W_{rel}^{i^{\prime},i})P_{i^{\prime}}], \label{Rate}%
\end{align}
with $P_{i}$ denoting the population probability of the SMM state $\left\vert
i\right\rangle $. The sequential rate $W_{\alpha}^{i,i^{\prime}}$ describes
the transition from the molecule state $\left\vert i\right\rangle $ to\textbf{
}$\left\vert i^{\prime}\right\rangle $ along with the electron tunneling into
or out of lead $\alpha$, and $W_{\alpha,\alpha^{\prime}}^{i,i^{\prime}}$
denotes the cotunneling rate from lead $\alpha$ to $\alpha^{\prime}$. The
above sequential and cotunneling rates can be calculated in terms of the
$T$-matrix with the help of generalized Fermi golden-rule
\cite{Xie,Bruus,Timm5,Timm6,Koch1,Koch2}. Moreover, the transition
selection-rules of spin relaxation are $\Delta n=0$ and $\Delta m=\pm1$, which
are similar to the cotunneling processes. The phenomenological spin-relaxation
rate from the state $\left\vert i\right\rangle $ to $\left\vert i^{\prime
}\right\rangle $ reads \cite{JBrel,Timmco}%
\begin{equation}
W_{rel}^{i,i^{\prime}}=\frac{1}{\tau_{rel}}\frac{1}{1+e^{(\varepsilon
_{i^{\prime}}-\varepsilon_{i})/k_{B}T}},
\end{equation}
where $\tau_{rel}$ is the relaxation time and $\varepsilon_{i}$ denotes the
energy of state $\left\vert i\right\rangle $. In addition, a dimensionless
parameter $\gamma=\hbar/(\Gamma\tau_{rel})$ called the spin-relaxation
strength is introduced to compare $\tau_{rel}$ with the sequential tunneling
time $\hbar/\Gamma$.

For the steady-state transport we have the stationary probabilities with the
condition $\frac{dP_{i}}{dt}=0$. Therefore, the stationary current
through\textbf{\ }the lead $\alpha$\ is given by%
\begin{align}
I_{\alpha}  &  =(-1)^{\delta_{\alpha L}}e\sum_{\alpha^{\prime}\neq
\alpha,ii^{\prime}}[(n_{i^{\prime}}-n_{i})W_{\alpha}^{i,i^{\prime}}%
P_{i}\nonumber\\
&  +(W_{\alpha,\alpha^{\prime}}^{i,i^{\prime}}-W_{\alpha^{\prime},\alpha
}^{i,i^{\prime}})P_{i}],
\end{align}
where the current flows from the left electrode to right one, and the
magnetization of SMM is $\left\langle S_{t}^{z}\right\rangle =\sum_{i}%
m_{i}P_{i}$. The TMR is defined as%
\begin{equation}
\text{TMR}=\frac{I_{P}-I_{AP}}{I_{AP}},
\end{equation}
where $I_{P}$ ($I_{AP})$ denotes the current through the SMM in the P (AP) configuration.

\section{Results and discussion}

We now analyze the effects of intrinsic spin-relaxation in both the sequential
and cotunneling regions with the assumption that the bias voltage is symmetric
at the SMM-electrode tunnel junction, i.e., $\mu_{L}=-\mu_{R}=V/2$. In the
following numerical calculations, the parameters are chosen as $S=2$,
$\varepsilon=0.5$\ meV, $|J|=0.2$\ meV, $U=1$\ meV, $K=0.05$\ meV,
$\Gamma=\Gamma_{L}=\Gamma_{R}=0.001$\ meV, $p_{L}=p_{R}=0.5$\ and
$k_{B}T=0.04$\ meV.

The TMR value indicated by color index as a function of the gate ($V_{g}$ )
and bias ($V$) voltages is plotted in Fig. 2. The electron transports are
dominated by cotunneling transitions in the voltage-regions\textbf{ }labeled
with the numbers $n=0$, $1$, $2$. While in the rest of regions the transports
are dominated by sequential processes. The left panels [(a)-(c)] of Fig. 2 are
referred to the results without the intrinsic spin-relaxation as a comparison,
and the corresponding TMR for the relaxation strength $\gamma=1$ is shown on
the right panels [(d)-(f)]. Comparing Figs. 2(a)-2(c) and Figs. 2(d)-2(f), we
can see the obvious variation of TMR induced by the intrinsic spin-relaxation,
which in most cases leads to the TMR suppression. However, when an external
magnetic field is applied along the easy-axis of SMM, the TMR can be increased
seen from the Figs. 2(c) and (f). In particular, a negative TMR value extends
from cotunneling to sequential regions.

\begin{figure}[pth]
\includegraphics[width=1\columnwidth]{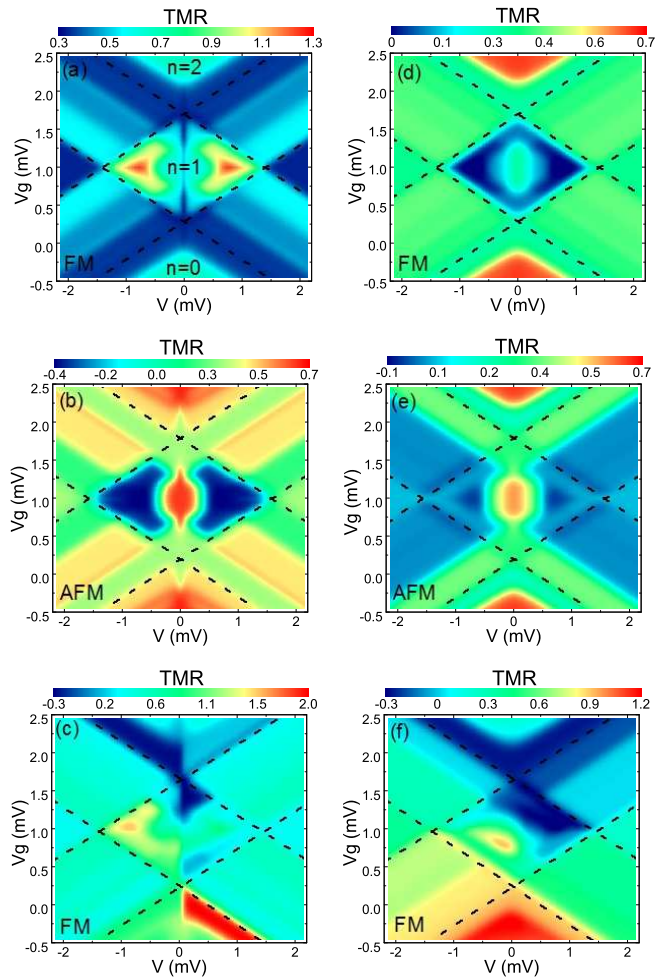}\caption{(Color online) TMR as
a function of the bias and gate voltages in the case of FM coupling [(a),(d)],
AFM coupling [(b),(e)], and FM coupling with a magnetic field along the
easy-axis of SMM [(c),(f)] for different relaxation strength $\gamma=0$ (left
panel) and $\gamma=1$ (right panel). The parameters are: $S=2$, $\varepsilon
=0.5$ meV, $|J|=0.2$ meV, $U=1$ meV, $K=0.05$ meV, $k_{B}T=0.04$ meV,
$p_{L}=p_{R}=0.5$, $\Gamma=\Gamma_{L}=\Gamma_{R}=0.001$ meV, and
$I_{0}=2e\Gamma/\hbar$.}%
\label{Fig:2}%
\end{figure}

\begin{figure}[pth]
\includegraphics[width=1\columnwidth]{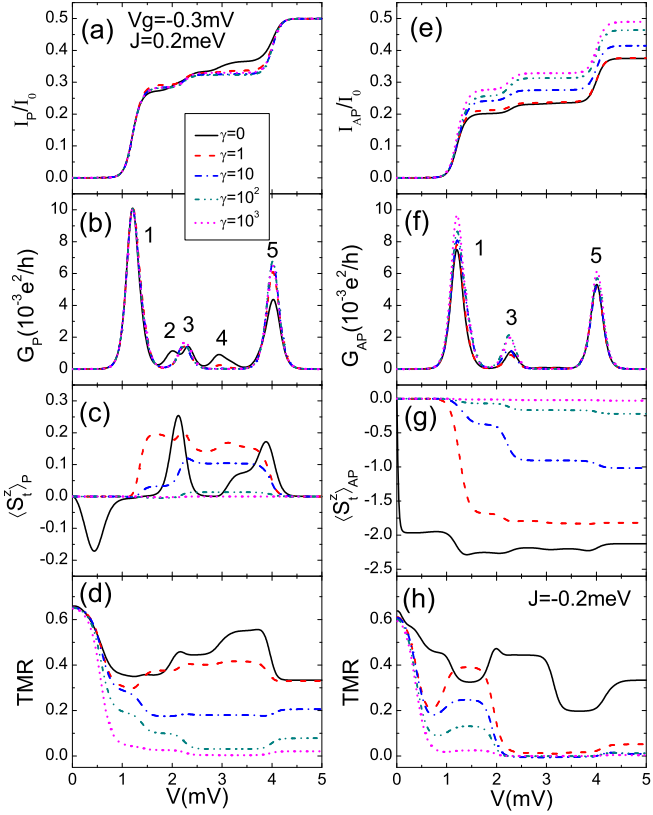}\caption{(Color online) The
bias voltage dependence of current $I$, differential conductance $G$, and the
magnetization of the SMM for the parallel [left panel (a)-(c)] and
antiparallel [right panel (e)-(g)] configurations with the gate voltage
$V_{g}=-0.3$ mV. TMR plots are shown in (d) and (h) for the FM and AFM
exchange couplings, respectively. }%
\label{Fig:3}%
\end{figure}

The bias voltage dependence of current, differential conductance,
magnetization of SMM, and TMR are shown in Fig. 3 for different relaxation
strength $\gamma$ with the gate voltage $V_{g}=-0.3$ mV, at which the ground
state of SMM is $\left\vert 0,2;\pm2\right\rangle $. For the P configuration
of FM leads, the conductance peak-$1$ [Fig. 3(b)], which corresponds to the
transition $\left\vert 0,2;\pm2\right\rangle \Leftrightarrow\left\vert
1,5/2;\pm5/2\right\rangle $, appears with the increase of bias voltage $V$ and
is not affected by the intrinsic spin-relaxation since the transition is
between the ground states of $n=0$ and $n=1$. When the initial state is
excited, the corresponding peak-$2$ ($\left\vert 0,2;\pm1\right\rangle
\Leftrightarrow\left\vert 1,3/2;\pm3/2\right\rangle $) and peak-$4$
($\left\vert 1,3/2;\pm3/2\right\rangle \Leftrightarrow\left\vert
2,2;\pm2\right\rangle $) disappear \cite{Timmco} in the existence of
spin-relaxation.\textbf{ }However, the peak-$3$ ($\left\vert 0,2;\pm
2\right\rangle \Leftrightarrow\left\vert 1,3/2;\pm3/2\right\rangle $) still
exists and especially the height of peak-$5$ ($\left\vert 1,5/2;\pm
5/2\right\rangle \Leftrightarrow\left\vert 2,2;\pm2\right\rangle $) increases
along with the population probability of the ground state $\left\vert
1,5/2;\pm5/2\right\rangle $. This enhancement phenomenon results from the
relaxation induced decay of excited states. In consequence, the current
$I_{P}$ between the peak-$1$ and peak-$3$ is slightly increased, while the
current between the peak-$3$ and peak-$5$ is decreased as shown in the Fig.
3(a). When the bias voltage is high enough such that all channels enter into
the transport window, the current $I_{P}$ remains in a constant value
independent of the relaxation. In addition, the magnetization of SMM vanishes
when the relaxation strength increases to $\gamma=10^{3}$ as shown in Fig.
3(c) (dotted line). The transport properties of AP configuration are quite
different from the P case. For example, in the absence of spin-relaxation the
magnetization of SMM [Fig. 3(g)] tends to a large negative value resulting
from the electron spin-flip process \cite{JBTMR1}, and the steady transport is
dominated by the negative-eigenvalue states of $S_{t}^{z}$. The
spin-relaxation, however, removes the above tendency and more
positive-eigenvalue states contribute to the transport. Therefore, the current
$I_{AP}$ [Fig. 3(e)] and differential conductance $G_{AP}$ [ Fig. 3(f)]
increase monotonously with the relaxation strength $\gamma$. The $\gamma
$-dependence of TMR spectrum for both cases of FM and AFM exchange couplings
are shown in Fig.3 (d) and (h), respectively. At the low bias voltage,
electron transport is dominated by elastic cotunneling processes via ground
states of $n=0$ and the TMR is essentially not affected by the relaxation.
With the increase of bias voltage, the inelastic cotunneling plays a role in
the transport and the current $I_{AP}$ increases due to the intrinsic
spin-relaxation. Therefore, the TMR almost vanishes at the large value of
$\gamma$. In the sequential tunneling region, the fast relaxation also leads
to vanishing TMR. In addition, the energy of state $\left\vert
1,3/2;m\right\rangle $ for the AFM exchange coupling is lower than that of the
state $\left\vert 1,5/2;m\right\rangle $, and the TMR decreases (with the
increase of relaxation strength) faster compared with the FM case.

\begin{figure}[pth]
\includegraphics[width=1\columnwidth]{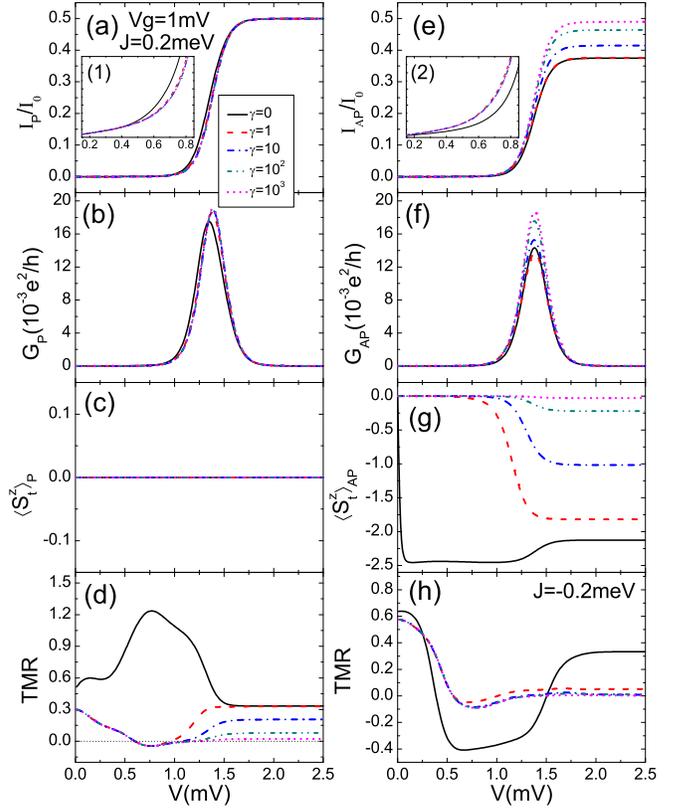}\caption{(Color online) The
bias voltage dependence of $I_{P}$ (a), $I_{AP}$ (e), $G_{P}$ (b), $G_{AP}$ (f),
$\left\langle S_{t}^{z}\right\rangle _{P}$ (c) $\left\langle S_{t}%
^{z}\right\rangle _{AP}$ (g), and TMR for the FM [(d)], AFM [(h)] exchange
couplings with the gate voltage $V_{g}=1$ mV. The cotunneling currents $I_{P}$
and $I_{AP}$ are shown in the insets of (a) and (e), respectively.}%
\label{Fig:4}%
\end{figure}

In Figs. 4(a)-4(c), we display the current, differential conductance and
magnetization of SMM in the P configuration as a function of the bias voltage
$V$ for the FM exchange coupling and $V_{g}=1$ mV. In this case, the gate
voltage corresponds to the electron-hole symmetry point, thus the
magnetization of SMM is always zero independent of the bias voltage $V$ [Fig.
4(c)] and the spectrum of differential conductance has only one peak seen from
Fig. 4(b). Moreover, the peak-position has a slight shift in the existence of
intrinsic spin-relaxation due to the increase of population probability of
state $\left\vert 1,5/2;\pm5/2\right\rangle $. For the P configuration, it can
be seen from Fig. 4(a) that $I_{P}$ is suppressed by the spin-relaxation in
both the cotunneling [the inset of Fig. 4(a)] and sequential tunneling
regions. But for the AP configuration, the current [Fig. 4(e)] and
differential conductance [Fig. 4(f)] are enhanced by the relaxation. In the
Fig. 4 (d), we show the bias voltage dependence of TMR with FM exchange
coupling for different relaxation strengths. The intrinsic spin-relaxation
leads to the decrease of TMR from a large positive to small negative value in
the cotunneling region. This can be understood from the behaviors of $I_{P}$
and $I_{AP}$ [see the insets of Figs. 4(a) and 4(e)]. For the AFM exchange
coupling, the TMR exhibits a different behavior from the FM case. At low bias
voltage, in which electron transport is dominated by cotunneling processes via
ground states $\left\vert 1,3/2;\pm3/2\right\rangle $, the intrinsic
spin-relaxation has a little effect on the TMR. When the bias voltage
increases, the enhancement of TMR induced by the spin-relaxation becomes
obvious. However, the large negative TMR almost disappears due to the fast
relaxation [Fig. 4(h)]. The above behavior of TMR for the AFM coupling is due
to the decrease of $I_{AP}$ in the cotunneling region. When all of transport
channels are open in the sequential region, the TMR of both FM and AFM
exchange couplings approaches zero when the relaxation strength increases to a
sufficiently large value.

Figure 5 shows the effects of the intrinsic spin-relaxation on the bias
voltage dependence of TMR in the existence of a magnetic field, which is
applied along the easy-axis of SMM. The longitudinal magnetic field lifts the
degeneracy of states $\left\vert n,S_{t};\pm m\right\rangle $ and the fast
spin-relaxation leads to the transition from the state $\left\vert
n,S_{t};m\right\rangle $ to the ground state $\left\vert n,S_{t}%
;S_{t}\right\rangle $ during the transport process. Moreover, this magnetic
field also leads to the symmetry-breaking with respect to the bias reversal in
the AP configuration, but the current remains symmetric in the P
configuration. The TMR is illustrated in Fig. 5(a) for $V_{g}=2.3$ mV, where
the transport is mainly through the ground state $\left\vert
2,2;2\right\rangle $ in the cotunneling region. It is found that the
spin-relaxation induced TMR-variation is negligibly small in the negative bias
voltage, while the spin-relaxation can lead to a great change of TMR varying
from positive to negative values in the positive bias voltage. Since the
currents $I_{P}$ for both positive and negative bias voltages are suppressed
equally by the spin-relaxation in the P configuration, the above asymmetric
behavior of TMR is determined by the transport in the AP configuration only.
For this case, there exists a competition between the spin-flip process
\cite{JBTMR1} (induced by the inelastic cotunneling) and the intrinsic
spin-relaxation. At the negative bias voltage, the above two mechanisms make
the SMM be trapped in the ground state $\left\vert 2,2;2\right\rangle $ and
the current is constant. However, at the positive bias voltage, the spin-flip
process makes the SMM tend to the state $\left\vert 2,2;-2\right\rangle $, but
the spin-relaxation drives it to the state $\left\vert 2,2;2\right\rangle $.
In consequence, the effect of relaxation can compensate the decrease of
$I_{AP}$ induced by the spin-flip. Moreover, the TMR for $V_{g}=1$ mV [Fig.
5(b)] and $V_{g}=-0.3$ mV [Fig. 5(c)] exhibits a similar behavior in the
cotunneling region.

\begin{figure}[pth]
\includegraphics[width=0.8\columnwidth]{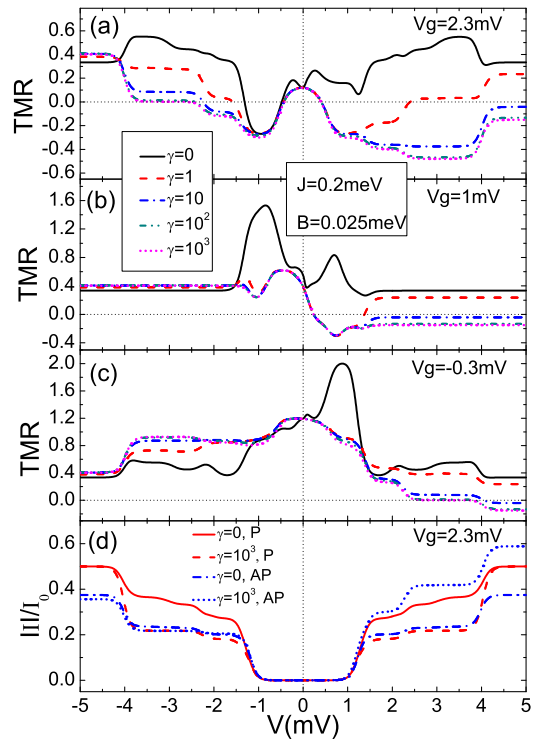}\caption{(Color online) TMR
as a function of bias voltage for different gate voltages (a) $V_{g}=2.3$ meV
(b)$1$ meV (c)$0.3$ meV. (d) Plots of absolute currents for different
relaxation strengths with a magnetic field along the easy-axis of SMM.}%
\label{Fig:5}%
\end{figure}

In order to understand the TMR behavior in the sequential region at
$V_{g}=2.3$ mV [see Fig. 5(a)], $I_{P}$ and $I_{AP}$ are plotted in Fig. 5(d)
for the relaxation strength $\gamma=0$ and $\gamma=10^{3}$. At the negative
bias voltage, both the spin-flip process \cite{JB,Timm2,Rossier,JBTMR1,Timm3}
and the intrinsic spin-relaxation can increase the population probabilities of
large positive-eigenvalue states of $S_{t}^{z}$ so that the relaxation induced
change of $I_{AP}$ is very small. However, at the positive bias voltage, the
current $I_{AP}$ is increased much by the relaxation. On the other hand, for
both positive and negative bias voltages in the P configuration, the
spin-relaxation results in the enhancement of sequential transition
$\left\vert 2,2;2\right\rangle \Leftrightarrow\left\vert
1,5/2;5/2\right\rangle $ for the spin-down electrons, which are minority in FM
electrodes, and thus $I_{P}$ decreases. As a result, the corresponding TMR can
vary from positive to negative values along with the increase of the
relaxation strength at a fixed bias voltage. In particular, a large negative
TMR is generated by the fast spin-relaxation at the positive bias voltage.
Fig. 5(c) shows the TMR as a function of the bias voltage for $V_{g}=-0.3$ mV,
where the ground state of SMM is $\left\vert 0,2;2\right\rangle $. Different
from the $V_{g}=2.3$ mV case, the TMR is enhanced by the spin-relaxation at
the negative bias voltage while it is suppressed even to zero at the positive
bias voltage. This is because that the intrinsic spin-relaxation leads to the
increase of $I_{P}$ by the enhancement of majority-electron transport through
the sequential transition $\left\vert 0,2;2\right\rangle \Leftrightarrow
\left\vert 1,5/2;5/2\right\rangle $ in the P configuration. In addition, when
the bias voltage is high enough such that all transport channels are open, the
$I_{AP}$ is slightly reduced by the spin-relaxation at the negative bias, but
it can be greatly enhanced at the positive bias. Therefore, the TMR is
slightly above the characteristic value of the sequential region in the
negative bias and becomes negative in the positive bias seen from the Figs. 5(a)-5(c).

\section{Conclusion}

In summary, the impact of the intrinsic spin-relaxation on electron transport
through a SMM with FM leads\ is investigated by the explicit calculation of
spin-relaxation dependence of transport quantities such as current,
differential conductance, magnetization and TMR in both the sequential and
cotunneling regions. The differential conductance peaks can be completely
suppressed by the spin-relaxation for the transition from initial
excited-states in the P configuration. The magnetization varies from a large
negative value to zero with the increase of spin-relaxation strength in the AP
configuration, and the current $I_{AP}$ is enhanced. Therefore, the TMR is
completely suppressed in the sequential region and can vary from a large
positive to slight negative value in the cotunneling region. When an external
magnetic field is applied along the easy-axis of SMM, the TMR is asymmetric
with respect to the bias reversal. For the cotunneling process, the TMR is
much more sensitive to the variation of spin-relaxation in positive bias than
in negative bias, especially, it can change from positive to negative values
in the former case. In the sequential region, a large negative TMR value can
be generated by the spin-relaxation in positive bias. Finally, in the high
bias limit the TMR at negative bias is slightly larger than the typical
sequential TMR value, while at positive bias it becomes negative due to the
spin-relaxation. It is expected that the above transport properties can be
observed experimentally in FM-SMM-FM spin valves \cite{Mugarza}.

\section{Acknowledgment}

This work was supported by National Natural Science Foundation of China (Grant
Nos. 11075099, 11275118, 11204203 and 11004124).

\end{document}